\documentstyle[11pt,paspconf,psfig]{article}

\begin{document}

\title{High Resolution Data on the Cloverleaf in the UV and in CO(7-6) :
a New Model of the Lens and a Model of the Molecular Region in the
Quasar at z=2.56 }

\author{D. Alloin}
\affil{CNRS-URA2052, SAp, CE Saclay, Gif-sur-Yvette, France.}

\author{Y. Mellier}
\affil{Institut d'Astrophysique de Paris and Observatoire de
Paris, DEMIRM, Paris, France.}

\author{J.-P. Kneib}
\affil{Observatoire Midi-Pyr\'en\'ees, CNRS-UMR5572, Toulouse, France.}





\begin{abstract}
We present a new investigation of the Cloverleaf (z$=$2.56) based on
the combination of archival HST/WFPC2 data, recent IRAM CO(7-6) maps
and wide field CFHT/FOCAM images.
The deep WFPC2 observation (F814W) shows a significant overdensity of
I$_{814W}\sim$ 23--25 galaxies around the Cloverleaf that we interpret
as the presence of a distant  cluster of galaxies along the line of sight.
 The Cloverleaf is probably
the result of the lensing effects of a system which includes,
in addition to a single galaxy,
one of the most distant clusters of galaxies ever detected.
We have modelled the lens accordingly. \\
The high-resolution IRAM/CO map provides for the first time
the orientation and the ellipticity of the CO spots induced by the
shear component.
Velocity - positional effects are detected at the 8$\sigma$ level
in the CO map. A strong limit can then be put on the size, shape
and location of the CO source around the quasar. The CO source is found
to form a disk- or ring-like structure orbiting the central engine at
$\sim$ 100km/s at a radial distance of $\sim$ 100pc, leading to
a central mass of $\sim$ 10$^{9}$ M$_\odot$ possibly in the form of a
black hole.
\end{abstract}




\section{Introduction}
The Cloverleaf is the gravitationally lensed image of the quasar
H1413+117 (14$^{\rm h}$ 15$^{\rm m}$ 46$^{\rm s}$.23;
$11^{\rm o}\ 29'\ 44''.0$ J2000.0) at
$z=2.558$ showing four spots with angular separations from $0''.77$ to
$1''.36$.  Since its discovery (Magain et al. 1988), the Cloverleaf has
been observed spectroscopically and imaged with ground based telescopes
in various bands from B to I as well as at 3.6 cm with the VLA. An early
model of the gravitational lens was derived by Kayser et al (1990). 
The main difficulty with this model lies in the fact that
it predicts, for a z=1.44 lens,
a mass of $\sim$ 2.5 $10^{11}$ $h^{-1}_{50}$ M$_\odot$
within  $0.7''$ (6 $h^{-1}_{50}$ kpc) radius, which would correspond to
a relatively bright normal galaxy:
so far, searches in the K band of the predicted `bright'
lensing galaxy have been unsuccessful (Lawrence 1996) and this fact
remains a mystery.\\
New data sets are available : the post-COSTAR HST/WFPC2 UV images (partly discussed in
Turnshek et al. 1997), maps and spectral informations 
in the molecular CO transitions (Barvainis et al
1994, Wilner et al 1995, Barvainis et al 1997, Yun et al 1997, Alloin et
al 1997) and CCD images of
the Cloverleaf over a 5' field of view (CFHT/FOCAM archive) allowing a
precise astrometry of the field obtained through different wavebands.  
New results  obtained from these data are discussed in
the following sections. 
Throughout the paper, we use H$_0$= 50 $h_{50}$ km/s/Mpc, $\Omega_0$=1
and $\Lambda=0$.

\begin{figure}
\vspace{1.75in}
\caption{Image of the Cloverleaf obtained with the IRAM telescope
at Plateau de Bure.
(a) is the total CLEANed image, (c) the CLEANed blue-shifted
image, (d) the CLEANed red-shifted
image and (b) the difference between the CLEANed red and CLEANed blue
image.
The CLEANed CO(7-6) maps were obtained with a natural beam of
$0.8''\times 0.4''$ at P.A. 15 deg. They have been restored with a
circular
$0.5''$ beam for comparison with HST data.
Contour spacing is 1.35 mJy/beam for 1a, 2 mJy/beam for 1c and 1d
and 3 mJy/beam for 1b, corresponding in each case to 2$\sigma$.
}
\label{fig:iram}
\end{figure}
\section{The IRAM interferometric CO(7-6) map}
Our previous CO(7-6) IRAM interferometer measurements (Alloin et al 1997)
have been complemented with observations at intermediate baselines.
The
combined data lead to the CLEANed integrated map
restored with a $0.5''$ circular beam shown in Figure~\ref{fig:iram}a.
In order to search for a velocity gradient, we have first
derived the spatially integrated line profile, following the procedure
 discussed in Alloin et al (1997). The new CO(7-6) line profile,
shown in Figure 2 of Kneib et al (1998), exhibits a marked asymmetry with
a
steep rise on its blue side and a slower decrease on its red side.
Excluding the central velocity channel (so that the split in velocity is
symmetric), we have built the blue
(-225,-25 km/s) and the red (+25,+225 km/s) maps displayed in
Figures~\ref{fig:iram}c and \ref{fig:iram}d respectively. The difference
between
the red-shifted and blue-shifted CLEANed maps (Figure~\ref{fig:iram}b)
establishes definitely the presence of a velocity gradient at the
8$\sigma$ level.\\
Measurements of the spot characteristics from the CO(7-6) image have
been performed (spot flux ratios, sizes and orientations) through a
fitting procedure in the visibility domain, as explained in Alloin et al
(1997).
The final parameters are provided in Table 1 of Kneib et al (1998) where the
spot sizes are intrinsic to the image, {\em i.e.} deconvolved by the
interferometer beam.
Although the measurements have not been corrected for seeing effects
(mean seeing estimate of the order of $0.2''$) the spots A, B and C
appear
to be definitely elongated (see Figure~\ref{fig:iram}b for spot
labels).

%

\section{HST/WFPC2 data set}
The HST data of the Cloverleaf have been
provided by the
ESO/ST-ECF Science Archive Facility (Garching).
Two data sets obtained with the WFPC2 are used here,
obtained by Turnshek in 1994 and by Westphal
 in 1995. Standard reduction procedures using IRAF/STSDAS
packages
have been applied. The absolute photometry was obtained using magnitude
zero-points given in Holtzmann et al. (1995).
Complete information about the final images 
 can be found in Kneib et
al (1998). \\
Regarding the relative astrometry and spot sizes, the
PC observations in filters F336W, F555W, F702W and F814W
provide similar results (see Table 1 in Kneib et al 1998).
The four spots are stellar-like with a FWHM $\approx$ 1.5
pixel or $0.068''$.
\\
The absolute photometry and the relative intensity ratios (Table
4  in Kneib et al 1998) have been computed with the Sextractor
software (Bertin \& Arnouts 1996). No PSF fitting was applied
as the four different spots are well separated on the PC.  The important
variation of
the intensity ratios in U compared to V, R and I band
could probably be  explained by absorption along the line of sight
by
intervening galaxies (HI clouds at redshift $\sim 1.7$ will act as very
efficient absorbers in the U band). Turnshek et al (1997) proposed that dust
extinction is the explanation with a preference
for an SMC-like dust extinction at the redshift of the quasar.

Figure~\ref{fig:mosaicI} shows the deep F814W image,
where the presence of numerous faint
objects over a $40''$ region in the environment of the Cloverleaf is
striking.
In order to quantify this effect we have computed the object number
density, $<$N$>$, and its dispersion, $\sigma_N$,
in the magnitude range I=23 to 25. We have estimated this density
in different regions across  the image: in a $40''$ diameter region
around the Cloverleaf where the
density contrast is clearly visible by eye and in various randomly
selected
areas of similar size.  On the whole frame we find a mean value
$<$N$>_{whole}=45 \pm 20$ objects/arcmin$^2$
(1 $\sigma$). The region around the Cloverleaf has $<$N$>_{Cl}=85 \pm 30
$
objects/arcmin$^2$, with a  peak at 130  objects/arcmin$^2$.  
 The bulk of the objects forming the
overdensity  are
red objects with $R_{702W}-I_{814W}\sim 0.9$ (from 0.7 to 1.2) and have
a small size ($\le 0.3"$).
While their morphology cannot be obtained from the HST images, their red
color suggest that we might be dealing with
E/S0 galaxies in a high-redshift cluster. This cluster could either be
linked to the quasar at redshift z$\sim$2.56 or
along the line of sight.
Hereafter, we {\em assume} this cluster to be at a redshift close
to that of the
narrow absorption systems observed by Turnshek et al (1988),
and  Magain et al (1988): either 1.438, 1.661, 1.87 or 2.07.
For the sake of simplicity, we shall adopt hereafter a value of 1.7.\\
Although the cluster candidate mentioned earlier is acting
as an additional lensing agent,
the small angular separation between the four spots implies that part of
the
lensing effect is produced by a galaxy located amid the four
spots of the quasar. The detection of the lensing galaxy would be of
great importance to model
accurately the mass distribution of the lens. Yet, its presence is still
elusive (Angonin et al. 1990; Lawrence 1996).
We tentatively used the deepest
WFPC2/F814W image in order to detect it. The
four
spots of
the quasar have been subtracted using the PSF model provided by a nearby star. We conclude that the
lensing galaxy must have $I>24$.  From modelling considerations (Bruzual
\& Charlot 1993), we  
find that 
the galaxy should have an absolute magnitude larger than $M=-21.3$ in
order not to be detected.  
 Hence, neither the visible nor the near
infrared data are  sufficient to constrain the mass of the lensing
galaxy.
But if we include  the likely presence of the distant
cluster, it is
no longer  necessary  to invoke a large mass for
the
lensing galaxy since a fraction of the convergence and the
deflection angle would be contributed by the cluster.

\section{Absolution registration of images in the  UV and the mm range}
In order to derive precisely the shear induced by the gravitational lens
on an extended source in the quasar, it is imperative to register with
a high accuracy the Cloverleaf image in a waveband corresponding to
  point-like images of the quasar (i.e. R-band corresponding to rest
wavelength 1967 \AA) and in a waveband corresponding to an extended
source in the quasar (i.e. the CO(7-6) molecular emission).  The high
precision required, better than $0.2''$, cannot be achieved from the HST
data because of their limited field of view. Hence, we used a set of R and I
bands for wide-field
exposures of the Cloverleaf field obtained under 0.6" seeing
conditions at CFHT by Angonin, Vanderriest and
Chatzichristou, with FOCAM. 
The absolute astrometry was performed using 6 reference stars across the
field, from the Cambridge APM database. This
allows  absolute positioning of the brightest spot A within an
accuracy of $0.15''$ (rms). Then, we have used the relative astrometry
of
spots B, C and D, with respect to A, from the HST image  (relative
accuracy $\pm 0.01''$). Finally, the IRAM CO(7-6) image is obtained
within an absolute
astrometric accuracy of $0.1''$ (rms). 
Therefore, it is possible to register both the optical and the
millimeter data in an absolute manner (Figure~\ref{fig:iram}), 
and within an accuracy of 0.15".
%
%
%
%
\begin{figure*}[t]
\caption{Image of the  HST field around the Cloverleaf.
 The four quasar spots are not resolved 
 but are located at the center of the WF3 chip (bottom-right).
The galaxies detected in the field 
 are overlayed with elliptical contours
indicating
their centroid, orientation and ellipticity. The white
contours are iso-number density of galaxies with $23<I<25$ (ranging from
30 to 80 galaxies/arcmin$^2$). A significant density
enhancement is clearly visible around the Cloverleaf, 
 suggesting the presence of a distant cluster of galaxies on 
its  line of sight.
 }
\label{fig:mosaicI}
\end{figure*}

\section{New model of the Cloverleaf gravitational lens}
The modelling of the Cloverleaf lens is based on the minimization
algorithm
described previously in several papers (Kneib et al 1993, 1996).
 The model
incorporates parameters of the lensing potential through a simple
analytical representation of the mass distribution, here a truncated
elliptical mass distribution (Kneib et al 1998).
The constraints used for the gravitational lens modelling are as
follows:
\begin{enumerate}
\item the {\sl relative positions} of the four quasar spots from the HST and
the
{\sl intensity ratio} taken in the R and I band, the least affected by
dust absorption.

\item the {\sl non-detection} of a 5th spot, which puts a limit on the
size of the lens core.

\item the {\sl position of the cluster center}  as measured from the
overdensity  of galaxies near the Cloverleaf.

\item both the cluster and the lensing galaxy are assumed 
to be at z=1.7.

\end{enumerate}

The {\sl relative intensity ratios and the measured shapes of the CO
spots} are used as a test of the model, and provide in addition information
on the size and geometry of the CO source in the quasar.

We have computed two types of model: model 1 which includes an
individual
galaxy with a dark halo at z=1.7 and model 2 which considers an
individual galaxy and a cluster both at z=1.7. These models
are not unique but give similar qualitative results.
The parameters for the lensing galaxy and the cluster component, as well
as the full details of the modelling can by found in Kneib et al
(1998).\\
In order to test, at first order, the lens model with the CO(7-6) map,
we have assumed the CO source to be elliptical
with a gaussian profile. We have fitted its position, size and
ellipticity so that it reproduces the observed CO image.
The upper limit of its size is provided by the CO elongated
spots A and B which are close to merging, but still clearly
separable.
Thus, the model must predict, in the image plane,
disconnected isocontours of the A and B spots. We found a typical size of
460$\times$230 pc (FWHM) for
model 1, and 155$\times$110 pc (FWHM) for model 2.
In the present case, $1''$ in the source plane translates into
$7.68h_{50}^{-1}$ kpc with the chosen cosmology.
A summary of the CO modelling is displayed in Figure~\ref{fig:modelco}.\\
The agreement between the
expected
amplifications and the observed UV flux ratios is relatively good.
The comparison between the expected amplification and the observed ones
in CO is not easy. Indeed, differential amplification due to the CO source
extent and location with respect to the diamond caustic can explain
the observed differences. The total amplification of the
CO emission is $\sim$ 18 for model 1 and $\sim$ 30 for model 2.
Following Barvainis et al (1997), these amplification factors
translate for model 1 (resp. model 2)
to molecular mass $M(H_2)$= 3. 10$^9$ M$_\odot$ (resp. 2.10$^9$
M$_\odot$) and $M(HI)$= 3. 10$^9$ M$_\odot$ (resp. 2.10$^9$ M$_\odot$).
These mass estimates are in good agreement with the dynamical mass
computed
in the next section, provided uncertainties in the inclination and
in the conversion from $M(CO)$ to $M(H_2)$.

\section{Constraints on the quasar CO source}
In addition to the size of the quasar CO source,
150--460 pc FWHM, as derived above, we can obtain some
information on its substructure using the velocity gradient observed in
the CO(7-6) line. The map
 corresponding to the blue side of the line
(-225,-25 km/s) shown in Figure~\ref{fig:iram}c shows that spot C is stronger and slightly displaced
inwards with respect to its counterpart in the map corresponding to the
 red part of the line (+25,+225 km/s), shown in Figure~\ref{fig:iram}d. A hint of this
effect can be found in Yun et al (1997) although a close comparison
of the IRAM results with the OVRO one is hampered by the fact that in
the OVRO study the spatial resolution is twice lower, the line profile
is not shown, the velocity interval considered -- 145 km/s -- is
narrower
than ours
-- 200km/s -- and not positioned precisely with respect to the line
center. Using the two lens models obtained previously, we optimized the
structure of the CO source to reproduce separately the red and the blue
images (Figure~\ref{fig:modelco}).\\
For model 1 [resp. model 2], we find a difference of
$0.03''$ ($\sim 230$ pc) [resp. $0.02''$ ($\sim 150$ pc)]
between
the center of regions emitting the blue and red parts of the CO line.
We have tested this procedure against uncertainties in the CO/HST images
registration. A change by $0.05''$ (half a CO map pixel) has a very
minor
impact on the positions of the source regions emitting the blue and red
parts of the line.
The quasar point-like visible source appears to be almost exactly
centered between the blue- and red-emitting regions.
This is reminiscent of a disk- or ring-like structure
orbiting the quasar at a radius of $\sim$ 100 pc [resp. 75 pc]
and with a Keplerian
velocity of $\sim$ 100 km/s (assuming a 90 deg inclination with respect
to
the plane of the sky), the resulting central mass would be
$\sim$ 10$^{9}$ M$_\odot$ [resp. $\sim$ 7.5 10$^{8}$ M$_\odot$].
Elaborating a more sophisticated (realistic) model of the CO source and
its link  with the BAL feature also observed in this quasar 
 should be performed in the future.
 Yet, such a geometry is consistent with  
 the molecular torus of the standard AGN model.
The resulting differential amplification  is likely 
at the origin of the asymmetry observed in the line profile
(Figure 2 in Kneib et al 1998).

\begin{figure*}

\begin{minipage}{4.5cm}
\end{minipage}
\begin{minipage}{4.5cm}
\end{minipage}
\begin{minipage}{4.5cm}
\end{minipage}
\begin{minipage}{4.5cm}
\end{minipage}
\begin{minipage}{4.5cm}
\end{minipage}
\begin{minipage}{4.5cm}
\end{minipage}
\begin{minipage}{4.5cm}
\end{minipage}
\begin{minipage}{4.5cm}
\end{minipage}
\begin{minipage}{4.5cm}
\end{minipage}
\begin{minipage}{4.5cm}
\end{minipage}
\begin{minipage}{4.5cm}
\end{minipage}
\begin{minipage}{4.5cm}
\end{minipage}

\caption{Results of the lens modelling of the Cloverleaf superimposed
on the HST image.
(a) is the
HST image overlaid with the CO observed (-225, +225 km/s);
(b) is the CO predicted for model 1, convolved by the interferometer
beam.
(c) is similar to (b) but for model 2.
(j) is the the CO  predicted for model 1, not convolved by the
interferometer beam.
(d), (e), (f) is similar to the first row but for the blue emission.
(g), (h), (i) is similar to the first row but for the red emission.
The dotted line is the corresponding critical line.
(k) gives the position of the best fitted CO sources for the blue
(dashed),
red (dotted) and total (solid) emission, for model 1.
(l) is similar to (k) but for model 2.
The central diamond-shape curve (in (b), (e), (h), (k) and (l))
is the internal caustic crossed by the lensed CO source at redshift
z=2.558.
}
\label{fig:modelco}
\end{figure*}

\section{Summary and outlooks}
This new analysis of the Cloverleaf reveals that
this is probably a complex lens which includes a lensing galaxy and an
additional distant lensing cluster of galaxies.
The reality of the cluster toward the Cloverleaf has still to be
confirmed independently.
Yet, most of the faint galaxies around the
Cloverleaf are found in the same magnitude and size ranges, as expected
if they indeed belonged to a cluster.  If the cluster is at a very
large distance, the shift of its galaxy
luminosity function up to higher apparent magnitude would  explain
why the number-density contrast of the cluster with respect to faint
field galaxies is lowered down to only a 4 $\sigma$ level. \\
This interpetation implies that the lensing galaxy may not be very
massive
and consequently may not be very luminous, helping to explain the
mystery
of the lensing galaxy not having been detected so far.   The
drawback is that, despite the constraint that the shapes of the
CO spots provide on the orientation of the mass density distribution,
it mandates sharing the
mass between the lensing-galaxy and the lensing-cluster which
 increases the number of possible lens configurations. 
 Further data 
 might  reveal the position and the shape of the light
distribution of the lensing galaxy.\\
Such information
would be useful to improve the mapping of the CO
source, as  emphasized by 
 Alloin et al. (1997). With the present-day data,
the CO source is found to be a disk- or ring-like structure with typical
radius
of $\sim$ 100 pc, under the lens composite model of a galaxy and a cluster at
z=1.7, leading to a central $\sim$ 10$^{9}$ M$_\odot$ object, typical
of a massive  black-hole.
It is amazing to see that a disk with such a
small intrinsic size can be spatially
``resolved" even at an angular distance as large as 1.6 $h^{-1}_{50}$
Gpc.\\
Although this remains to be confirmed independently, the discovery of a
distant cluster of galaxies on the line of sight to
the Cloverleaf is remarkable because it reinforces the suspicion that
many bright
high redshift quasars are magnified by cluster-like systems at large
distances. This was already reported from analyses in the fields of the
doubly imaged quasar Q2345+007 (Bonnet et al
1993; Mellier et al 1994; van Waerbeke et al 1997),
where a cluster candidate is expected to be at z$\sim$ 0.75
(Pell\'o et al 1996) and of MG2016 where the X-ray emission
of the intra-cluster gas has been observed and for which the Iron
line (from X-ray spectroscopy) gives a redshift z$\sim$1
(Hattori et al 1997). 

\acknowledgments
We thank  M.-C. Angonin and C. Vanderriest
for providing their CFHT images. We thank J. B\'ezecourt, S. Charlot,
J.-M. Miralles and  R. Pell\'o, for useful discussions. 


%
%

%

\end{document}